\documentclass[aps,prl,twocolumn,showpacs,amsmath,superscriptaddress,a4paper,amssymb,10pt]{revtex4-1}
\usepackage{graphicx}
\usepackage{CJK}
\usepackage{epstopdf}
\usepackage{amsmath,amssymb}
\usepackage{xcolor}
\usepackage[linkcolor=violet,citecolor=blue,colorlinks=true,urlcolor=purple]{hyperref}

\definecolor{rvsdcolor}{RGB}{30,95,20}

\newcommand{\rev}[1]{{\color{rvsdcolor} #1}}

\def\h2p{H$_2^+$}

\begin{document}
\begin{CJK*}{UTF8}{gbsn}
\title{Zero-energy proton dissociation of H$_2^+$ through stimulated Raman scattering}

\author{Xinhua Xie (谢新华)}
\email[Electronic address: ]{xinhua.xie@tuwien.ac.at}
\affiliation{Photonics Institute, Technische Universit\"at Wien, A-1040 Vienna, Austria, EU}
\author{Stefan Roither}
\author{Seyedreza Larimian}
\author{Sonia Erattupuzha}
\author{Li Zhang (张丽)}
\author{Daniil Kartashov}
\affiliation{Photonics Institute, Technische Universit\"at Wien, A-1040 Vienna, Austria, EU}
\author{Feng He (何峰)}
\affiliation{Key Laboratory for Laser Plasmas (Ministry of Education) and Department of Physics, Shanghai Jiao Tong University, Shanghai 200240, People's Republic of China}
\author{Andrius Baltu\v{s}ka}
\author{Markus Kitzler}
\email[Electronic address: ]{markus.kitzler@tuwien.ac.at}
\affiliation{Photonics Institute, Technische Universit\"at Wien, A-1040 Vienna, Austria, EU}

\date{\today}

\begin{abstract}
We show that (near-)zero energy proton emission from \h2p in strong two-color and broadband laser fields is dominated by a stimulated Raman scattering process taking place on the electronic ground state. 
It is furthermore shown that in the (near-)zero energy region the asymmetry in proton ejection induced by asymmetric laser fields is due to the interplay of several processes, rather than only pathway interferences, with vibrational trapping (or bond-hardening) taking a key role. 
\end{abstract}

\maketitle

\end{CJK*}

Much of our understanding used in the research on manipulating molecular reactions  with strong, tailored light fields \cite{Kling2013, Xie2012_CE, Alnaser2014_acetylene, Kubel2016, Alnaser2017} 
is footed on work done on H$_2$ \cite{Sheehy1995, Thompson1997, Kling2006, Kremer2009, Fischer2010, Znakovskaya2012, Kling2013a, Xu2013, Xu2014,  Ray2009, Xu2016a, Wanie2016}.
Key concepts emerging from the research on H$_2$ are, e.g.,  
the emergence of light-induced molecular potentials (LIPs) \cite{Giusti-Suzor1995, Posthumus2004},
bond-softening via the net-absorption of one \cite{Bucksbaum1990, Zavriyev1990}, 
two \cite{Zavriyev1990, Giusti-Suzor1990} 
or more \cite{McKenna2008, McKenna2012a} photons, 
or bond-hardening \cite{Giusti-Suzor1992, Zavriyev1993, Frasinski1999, Posthumus2000, Moser2009}, also known as molecular stabilization or vibrational trapping (VT);
%
%
see the reviews Refs.\,\cite{Giusti-Suzor1995, Posthumus2004, Ibrahim2018, Li2017} for further details. 
%
However, even for this simplest of all molecules there exist still a number of
issues awaiting clarification,
in particular in the family of bond-hardening phenomena.
Examples include direct experimental confirmation of light-induced conical intersections (LICIs) \cite{Sindelka2011, Halasz2011, Badanko2016, Natan2016}, or a generally accepted picture of the concept of trapping that has been challenged by McKenna~et~al. \cite{McKenna2009a}.



\begin{figure}[bt]
\centering
\includegraphics[width=0.5\textwidth,angle=0]{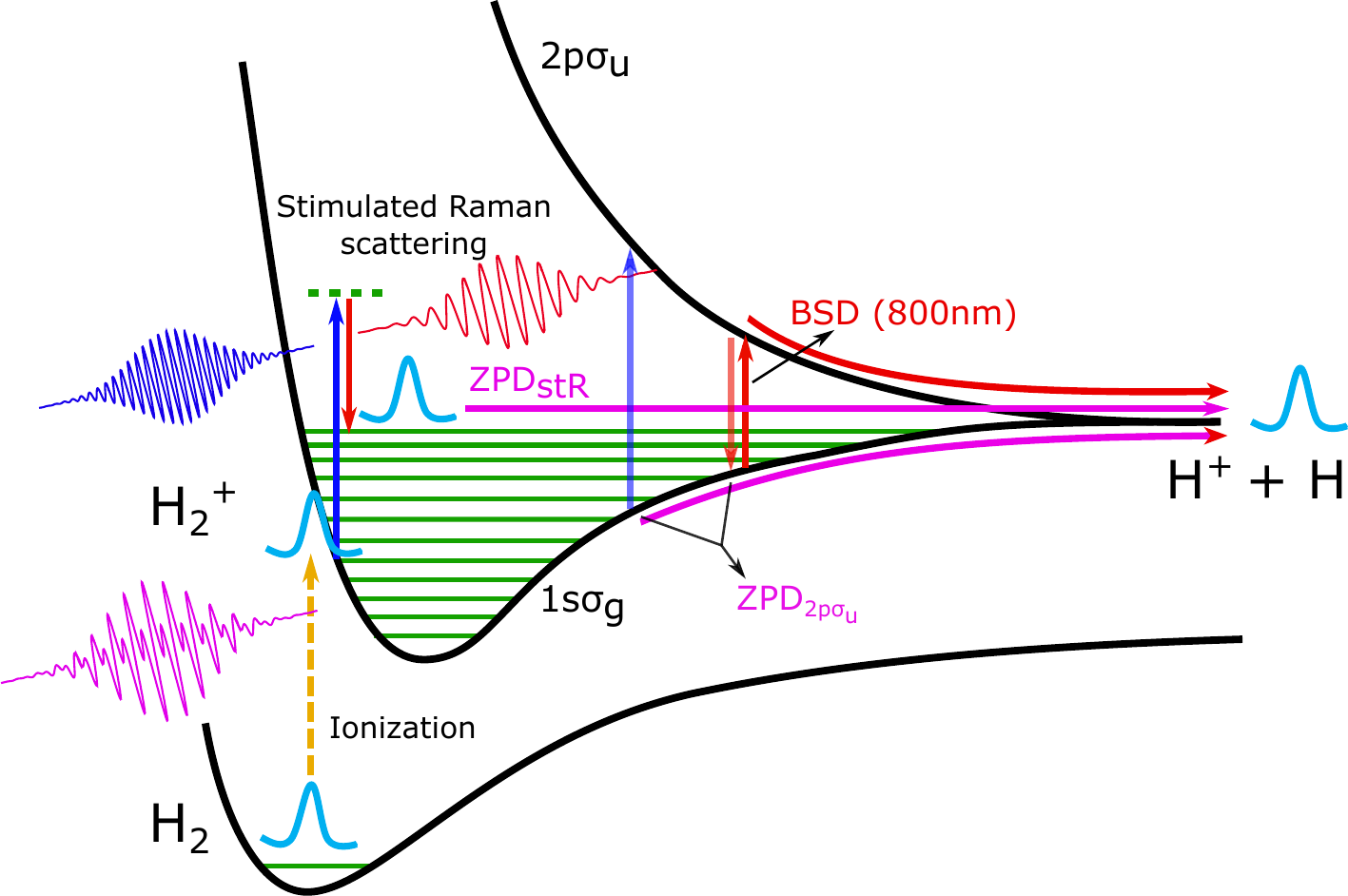}
\caption{Schematic of dissociation processes of H$_2$ relevant for the production of protons with near-zero energy in a two-color laser field (pink waveform in the lower left). Blue and red arrows indicate photons with 400-nm and 800-nm wavelength, respectively. ZPD and BSD denote zero photon dissociation and bond-softening dissociation, respectively. ZPD$_{\textrm{stR}}$ indicates ZPD via a stimulated Raman scattering process proceeding only on  the 1s$\sigma_g$ potential energy curve of H$_2^+$, and ZPD$_{\textrm{2p}\sigma_u}$ denotes ZPD by transient population of the 2p$\sigma_u$ curve via absorption of a 400-nm photon and later emission of a 800-nm photon.
} \label{fig:peda}
\end{figure}

Here, we focus on a bond-hardening process that leads to 
protons with (near-)zero kinetic energy during the dissociation H$^+_2 \rightarrow \textrm{H}^+ + \textrm{H}$. 
This dissociation pathway 
has been predicted \cite{Giusti-Suzor1992} and observed \cite{Posthumus2000} decades ago. 
It has been explained as bond-hardening at the zero-photon crossing of the Floquet ladder 
through dissociation involving the net-absorption of zero photons (zero photon dissociation, ZPD) \cite{Giusti-Suzor1992, Frasinski1999, Posthumus2000}.
However, because during ionization of H$_2$ at the Franck-Condon region the probability for populating vibrational levels higher than $\nu = 5$ is small, it necessitates laser wavelengths $\lesssim 400$\,nm to efficiently drive this process \cite{Giusti-Suzor1992, Posthumus2000}. 
%
%
\rev{Several recent experiments have shown that the yield of protons with (near-)zero energy increases particularly strongly when two-color laser fields are used to drive the H$_2$ dissociation process, e.g., Refs.~\cite{Moser2009, Ray2009, Gong2014c, Xie2016, Xu2016a, Zhang2017}. But also in experiments \cite{Kling2013a, Xu2013, Gaire2011} and simulations \cite{McKenna2008, McKenna2012a} applying few-cycle pulses with a broadband spectrum centered around $730-750$\,nm a notable enhancement of the \mbox{(near-)}zero energy proton yield was observed. }
\rev{The appearance of these low-energy protons in two-color fields was explained by a two-step process, where after ionization a 400-nm photon is resonantly absorbed at a stretched H-H$^+$ bond to transiently populate the ${\textrm{2p}\sigma_u}$ state and, subsequently, at a still further stretched bond, a 800-nm photon is emitted, returning the population to the $1s\sigma_g$ state where H$_2^+$ finally dissociates via the net-absorption of zero photons (ZPD), see, e.g., Refs.~\cite{Ray2009, Moser2009, Gong2014c, Zhang2017}. 
To stress the involvement of the 2p$\sigma_u$ state in this ZPD-process, we will refer to it as ZPD$_{\textrm{2p}\sigma_u}$, see Fig.~\ref{fig:peda} for a visualization.}
%

In this Letter we show experimentally that the yield-enhancement of protons with (near-)zero energy observed in two-color fields \cite{Moser2009, Ray2009, Gong2014c, Xie2016, Xu2016a, Zhang2017} and, with a somewhat smaller probability, also in broadband few-cycle pulses \cite{Kling2013a, Xu2013, Gaire2011, McKenna2008, McKenna2012a} is dominantly caused by a stimulated Raman scattering process, denoted by ZPD$_\textrm{stR}$ in Fig.~\ref{fig:peda}, rather than by the ZPD$_{\textrm{2p}\sigma_u}$ process.
%
%
Our work furthermore outlines the connection between the stimulated Raman scattering process and other processes leading to low-energy protons indicated in Fig.~\ref{fig:peda}, thereby filling the gaps of our thus far incomplete understanding of H$_2$-dissociation in this energy range.

\rev{In our experiments we employed broadband 5-fs pulses [center wavelength (CL) $740$\,nm] as well as narrow-band 25-fs pulses [CL $800$\,nm] and their frequency doubles [CL $400$\,nm, duration $50$\,fs].} With the narrow-band pulses we also generated two-color fields $E(t)={E}_{\omega_{800}} \cos(\omega_{800} t) + {E}_{\omega_{400}}\cos(\omega_{400} t + \Delta \varphi)$ with ${E}_{\omega_{800}}\approx{E}_{\omega_{400}}$ in the focus. All pulses were polarized along $z$. 
The relative phase $\Delta \varphi$ was varied 
using a glass wedge pair.
We used a reaction microscope \cite{Doerner2000, Ullrich2003} to measure the three-dimensional momentum vectors of electrons and ions emerging from the interaction of H$_2$ molecules with the two-color fields. 
The laser beam was focused onto an ultrasonic jet of H$_2$ in the interaction chamber (background pressure $1.3\times10^{-10}$\,mbar). 
Electrons and ions were guided by weak magnetic (6.4\,G) and electric (2.5\,V/cm) fields along the spectrometer axis ($z$-direction) to two position and time sensitive multi-hit detectors situated at opposite ends of the interaction chamber.
More details on our experimental setup can be found in our previous publications \cite{Xie2012_CE, Zhang2014a, Xie2016, Xie2017b}.

\begin{figure}[t]
\centering
\includegraphics[width=0.45\textwidth,angle=0]{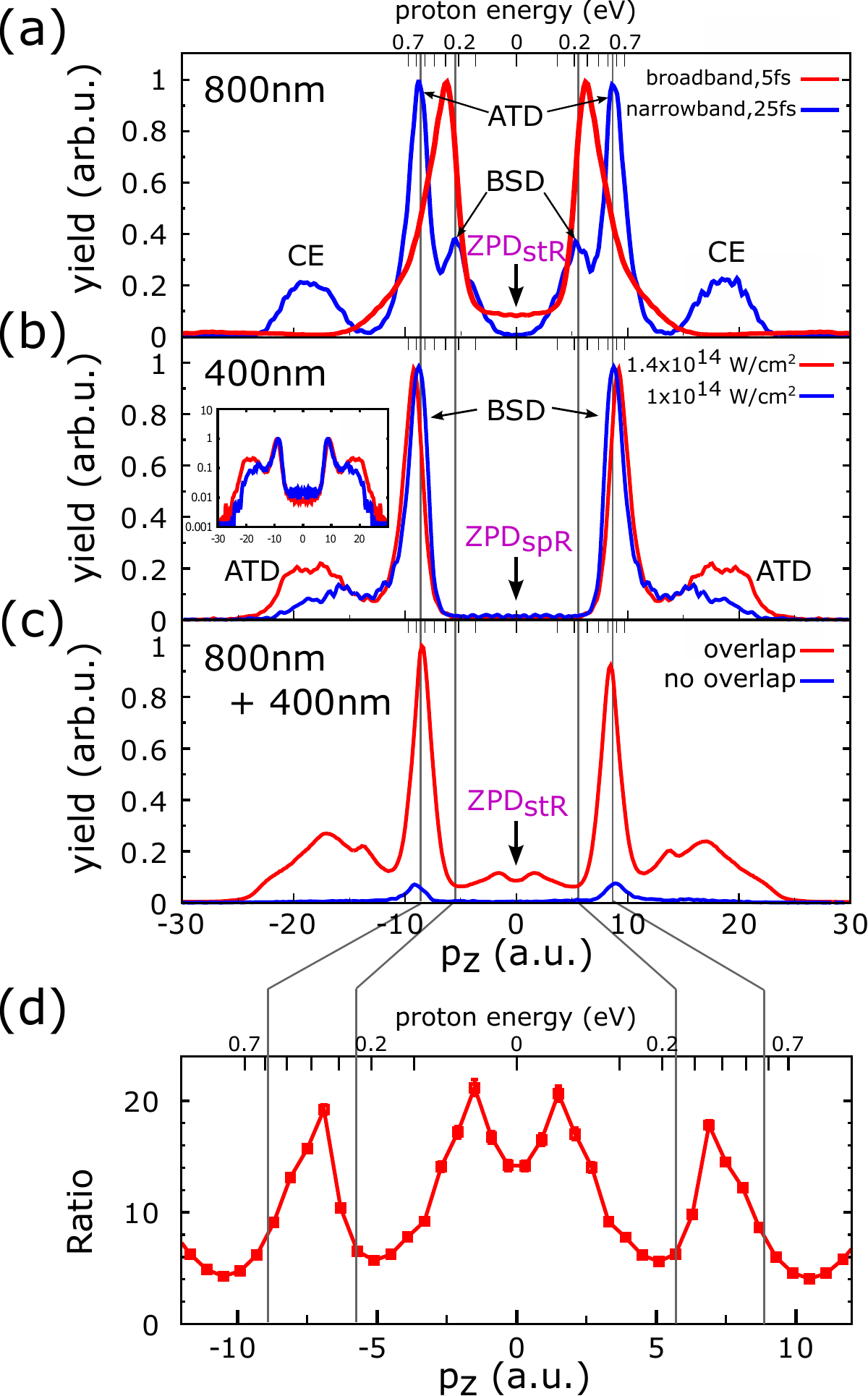}
\caption{Momentum distributions of protons along the laser polarization direction $z$. \rev{(a) Distributions measured with narrow-band pulses [FWHM bandwidth (BW) $50$\,nm around 800-nm, duration $25$\,fs] (blue line), in comparison with broadband laser pulses [FWHM BW roughly $300$\,nm around 740\,nm, duration $5$\,fs] (red line), normalized to maximum. Intensities of both pulses $2\times 10^{14}$\,W/cm$^2$.} CE: coulomb explosion. (b) Distributions measured with narrow-band 400-nm pulses generated by frequency doubling  (normalized, pulse peak intensities encoded by colors). Inset: same distributions on logarithmic scale. 
(c) Distributions measured with two-color pulses (800\,nm + 400\,nm); peak intensity 1$\times10^{14}$ W/cm$^2$ for each color. 
(d) Ratio of yields taken from (c) between the cases with and without temporal overlap of the two pulses, normalized to the yield of \h2p. 
} \label{fig:pz}
\end{figure}

Measured momentum distributions of protons observed along the laser polarization direction $z$ observed with different pulses are shown in Fig.~\ref{fig:pz}. 
For the single-color measurements [Figs.~\ref{fig:pz}(a) and (b)] we clearly identify the well-known peaks associated with the dissociation at the one-photon and the two-photon crossings \cite{Bucksbaum1990, Zavriyev1990, Frasinski1999}, usually called 
bond-softening dissociation (BSD) and above threshold dissociation (ATD)   \cite{Ibrahim2018, Li2017}, cf. the labeling in the figures. 

The momentum distributions in Figs.\,\ref{fig:pz}(a) and (b) show that there are almost no protons observed with momenta smaller than 4\,a.u. 
for both the narrow-band 800-nm and 400-nm pulses alone. 
In contrast, Fig.~\ref{fig:pz}(c) shows that if both pulses are overlapped in time and space the yield of protons with very small energy dramatically  increases.
A similar but somewhat less pronounced increase is also observed when broad-band pulses are used [Fig.~\ref{fig:pz}(a)].
Evidently, the dissociation process leading to these abundant near-zero-energy protons involves photons of distinctively different colors. This is clearly confirmed by the absence of the near-zero-energy protons in the narrow-band pulses and, even more clearly, by a cross-check measurement where the 800-nm and 400-nm pulses were applied with a time-delay of about 100\,fs such that both colors are supplied temporally separated [Fig.~\ref{fig:pz}(c)]. 
%
%
What is the mechanism behind the appearance of the near-zero-energy protons in the two-color and broad-band pulses?

To answer this question, let us discuss the possible dissociation pathways that can lead to protons with energy close to zero, cf. Fig.~\ref{fig:peda}.  
Dissociation starts after the ionization step H$_2 \rightarrow \textrm{H}_2^+ + e^-$. 
Following the Franck-Condon principle, vibrational states around $\nu$=5 will be dominantly populated in \h2p during ionization \cite{McKenna2012a}.
The dissociation barrier for $\nu$=5 is about 1.6\,eV. 
Thus, one 800-nm photon is not sufficient to populate vibrational states near the dissociation threshold.
Contrarily, the energy of one 400-nm photon is large enough to populate such vibrational states and to cause dissociation via a ZPD process, in which a 400-nm photon is absorbed and a photon with lower energy is emitted by spontaneous Raman scattering \cite{Posthumus2000}. We abbreviate this zero-photon dissociation process by ZPD$_\textrm{spR}$.
%
%
However, the probability of the spontaneous Raman scattering process is notoriously small. 
The inset of Fig.~\ref{fig:pz}(b) shows 
that for the 400-nm pulse we observe an accordingly small amount of protons with near-zero energy.

To explain the strong enhancement of the near-zero-energy proton yield in the two-color field [Fig.~\ref{fig:pz}(c)], we propose that a \textit{stimulated} Raman scattering process is at work. In this process, a high-lying vibrational state near the dissociation threshold is populated by the absorption of one 400-nm photon and the emission of one 800-nm photon, see the illustration in Fig.~\ref{fig:peda}. 
Equivalent to the ZPD$_\textrm{spR}$ mechanism, in this process also a net-amount of zero photons is absorbed. However, the lower-energy photon is supplied by the second wavelength in the two-color pulse and, thus, this ZPD process becomes stimulated. We denote it by ZPD$_\textrm{stR}$. 
The strong enhancement of the (near-)zero energy proton yield in the two-color field shown in Fig.~\ref{fig:pz}(c) is therefore explained by the much higher cross-section of the stimulated Raman process as compared to the spontaneous Raman scattering process.
\rev{Analogously, we also ascribe the increase of the (near-)zero energy proton yield observed with the broad-band pulse [Fig.~\ref{fig:pz}(a)] to ZPD$_\textrm{stR}$ with photons from the red and blue wings of the spectrum. Although the photon energy difference from the red and blue wings is not sufficient to completely reach the dissociation threshold directly from $\nu=5$, the stimulated Raman process can still take place from higher vibrational states. As  these states are populated less probably during ionization, the yield-enhancement for the very small (near-)zero energy region is less pronounced than for the still more broadband two-color field.  Nevertheless, this enhancement at the very low energies is a clear sign of the action of the ZPD$_\textrm{stR}$, in accord with interpretatons given in earlier work \cite{Gaire2011}.}

The ZPD$_\textrm{stR}$ mechanism is fundamentally different from the ZPD$_{\textrm{2p}\sigma_u}$-mechanism described in, e.g., Refs.~\cite{Ray2009, Moser2009, Gong2014c, Zhang2017} and outlined  above, cf. Fig.~\ref{fig:peda}. 
%
The ZPD$_{\textrm{2p}\sigma_u}$-mechanism requires a transition from the 1s$\sigma_g$ state to the 2p$\sigma_u$ state of H$_2^+$. 
As the transition probabilities are largest when the two photons are resonant with these two states, this mechanism necessitates that the two photons are absorbed respectively emitted at two different internuclear distances. Therefore, it inevitably implies the involvement of nuclear motion and a delay between the absorption and emission steps.
In contrast, the ZPD$_\textrm{stR}$ mechanism only involves the 1s$\sigma_g$ state and may happen directly within the Franck-Condon region without any nuclear motion.

The two processes, ZPD$_\textrm{stR}$ and ZPD$_{\textrm{2p}\sigma_u}$, generate protons in slightly different kinetic energy ranges. Because ZPD$_\textrm{stR}$ can, starting from around $\nu=5$, reach the dissociation threshold,  
the kinetic energy of the protons can reach down to zero. ZPD$_{\textrm{2p}\sigma_u}$, in contrast, can only take place from higher vibrational levels
that enable reaching internuclear distances where the 800-nm BSD process becomes available. 
As a result, ZPD$_{\textrm{2p}\sigma_u}$ leads to somewhat higher proton energies.  
\rev{Simulations and coincidence measurements performed in Ref.~\cite{Zhang2017} show that the yield of protons produced by ZPD$_{\textrm{2p}\sigma_u}$ peaks around $100$\,meV and becomes negligibly small below $30$\,meV. 
This leveling off at this proton energy can be explained by the finite bandwidths of the laser pulses which inhibit a larger spread around the peak proton energy of $100$\,meV down to smaller energy values.
}
%
\rev{In contrast, the ZPD$_\textrm{stR}$ process can populate vibrational levels   down to the dissociation threshold for both the two-color and broadband pulse. Even though in the latter case the process needs to start from higher $\nu$ (as explained above), the dissociation threshold is still reachable due to the smaller energies of the blue spectral portion. 
Thus, the enhanced yield at (near-)zero energies visible in Fig.~\ref{fig:pz}(a) is clear evidence for the action of the stimulated Raman process.}

\rev{Further evidence is obtained from the normalized ratio of the measured proton yields with and without overlap of the 800-nm and 400-nm  pulses, shown in Fig.~\ref{fig:pz}(d).}
Significant enhancement and suppression of the relative yields is observed at distinct values of the proton momentum. 
The enhancement around 7\,a.u. ($\approx 400$\,meV) and the suppression at 10\,a.u. ($\approx 700$\,meV) originate from the  fact that in the two-color field dissociation via the absorption of three 800-nm photons and the emission of one 400-nm photon becomes possible \cite{Ray2009}. 
These processes are not the primary subject of the present discussion. 
We are interested in the features at smaller momenta $|p_z| \lesssim 5$\,a.u. ($\lesssim 180$\,meV). 
Take the dip at 5\,a.u. This feature constitutes indirect evidence for the ZPD$_\textrm{stR}$ process: Since ZPD$_\textrm{stR}$ can happen directly in the Franck-Condon region, it depopulates the nuclear wave packet before it moves further along on the 1s$\sigma_g$ state to reach the internuclear distance where BSD of 800-nm takes place. As a result, the 800-nm BSD process becomes suppressed resulting in the dip at 5\,a.u.
Direct evidence for the ZPD$_\textrm{stR}$ process can be seen at $|p_z| \lesssim 3$\,a.u. ($\lesssim 70$\,meV):
As discussed above, the contributions from ZPD$_{\textrm{2p}\sigma_u}$ in this proton energy range are negligibly small \cite{Zhang2017} and only ZPD due to a Raman process can explain such low-energy protons \cite{Bucksbaum1990}. 
Thus, the huge yield-enhancement in comparison with the ZPD$_\textrm{spR}$ process of the single-color 400-nm pulse shown in Fig.~\ref{fig:pz}(d) is
clear evidence for the ZPD$_\textrm{stR}$ process.
%



\begin{figure}[t]
\centering
\includegraphics[width=0.45\textwidth,angle=0]{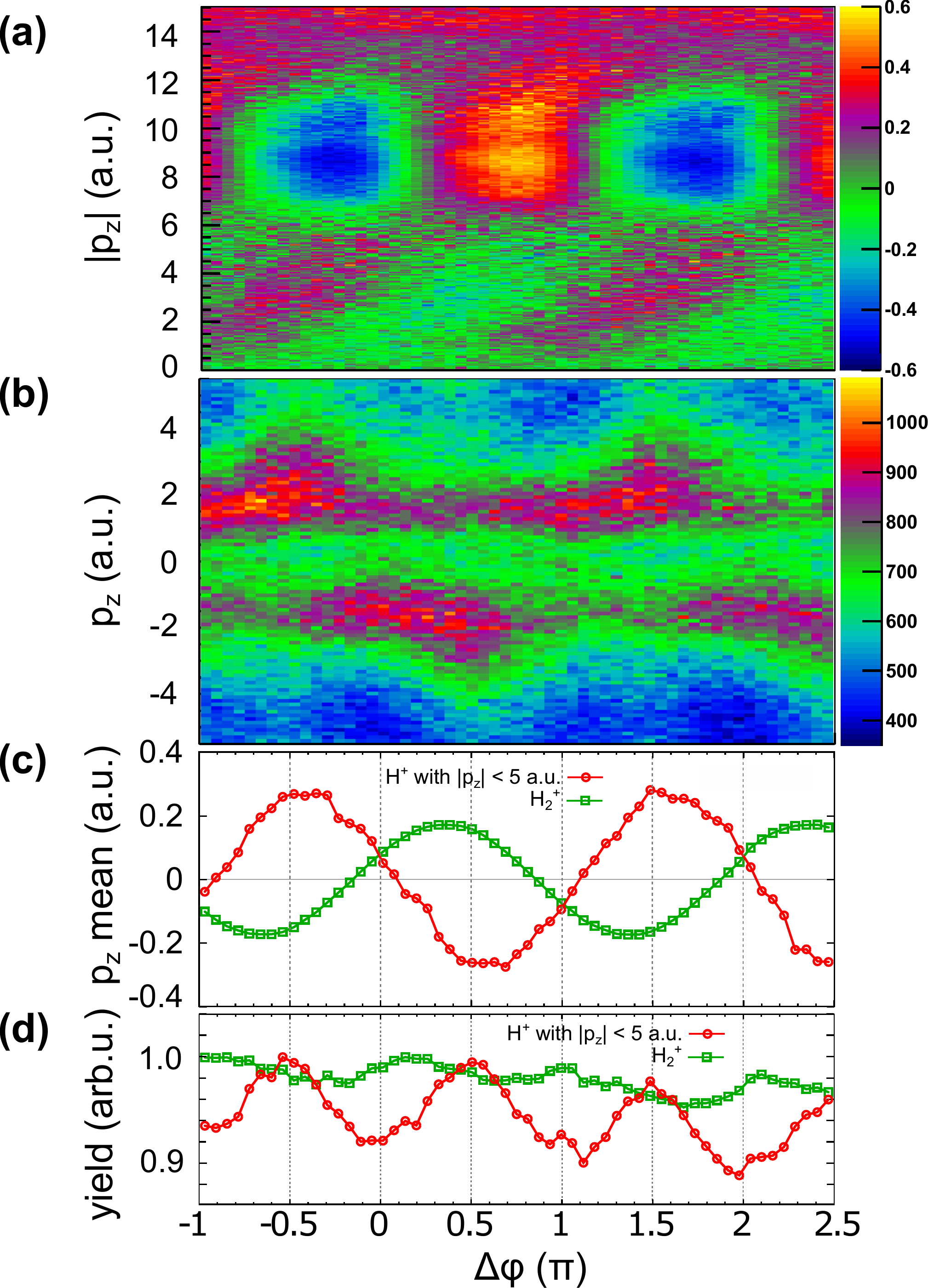}
\caption{(a) Asymmetry of proton emission (as defined in the text) as a function of $|p_z|$ over the relative phase $\Delta \varphi$ between 800-nm and 400-nm pulses. 
(b) Proton momentum distributions in the low-momentum region over $\Delta \varphi$.
(c) Mean momentum values over $\Delta \varphi$ calculated for the distributions in  (b) (red circles) and for H$_2^+$ (green squares). 
(d) Yields of protons from (b) (red circles) and for \h2p over $\Delta \varphi$, both normalized at their respective maxima.} \label{fig:pz_asym_phase}
\end{figure}

Dissociation of H$_2$ in two-color fields may lead to notable $\Delta \varphi$-dependent up-down-asymmetries in the proton yield, $A=(P_+-P_-)/(P_++P_-)$, with $P_+$ the yield of protons ejected upwards ($p_z>0$) and $P_-$ the downwards proton yield ($p_z<0$), as has been observed in many experiments, e.g., Refs.~\cite{Sheehy1995, Thompson1997, Ray2009, Xu2016a, Kling2013, Wu2013, Alnaser2017}. 
The usual explanation for the asymmetry in the low-energy region is wavepacket interference 
between dissociation on the 2p$\sigma_u$ state (due to 800-nm BSD) and dissociation on the 1s$\sigma_g$ state (due the ZPD$_{\textrm{2p}\sigma_u}$). 
\rev{Now, having established that protons below about 30\,meV are ejected dominantly along the ZPD$_\textrm{stR}$ pathway while the two other pathways are significantly weaker, one should wonder about the origin of the asymmetry in this energy range. If one relatively stronger pathway interferes with two weaker ones, the result is not easily predictable. 
Indeed, we measure significantly smaller values for $A(\Delta \varphi, |p_z|<5)$ as compared to $A(\Delta \varphi, |p_z|>5)$, where more pathways are open, see Fig.~\ref{fig:pz_asym_phase}(a). 
}

\rev{To understand how the asymmetry pattern $A(\Delta \varphi, |p_z|<5)$ is created 
one needs to look into the details of proton ejection in this momentum range. As we will show, it is governed by the interplay of several processes.
Fig.~\ref{fig:pz_asym_phase}(b) shows the momentum distributions of the protons as a function of $\Delta \varphi$. 
Two features are apparent: 
Their mean values, $\bar{p}_z$, vary periodically with $\Delta \varphi$, and there is a $\Delta \varphi$-independent trench visible for $|p_z| \lesssim 1$\,a.u.
Obviously, the $\Delta \varphi$-oscillation of the spectra [$\bar{p}_z(\Delta \varphi)$ is shown by red circles in Fig.~\ref{fig:pz_asym_phase}(c)] and their overlap with the trench is responsible for the observed asymmetry $A(\Delta \varphi, |p_z|<5)$, as the trench eats away the low-momentum parts of the spectra. 
}
%
%
One reason for the variation of $\bar{p}_z$ could be the center-of-mass (CM) recoil momentum that is imparted to H$_2^+$ during the ionization step, according to $p_z^{\textrm{H}_2^+} = p_z^\textrm{CM}(\Delta \varphi)=A_z(t_i, \Delta \varphi)$, where $A_z$ is the laser vector potential along $z$-direction and $t_i$ the instant of ionization. In a two-color field, $p_z^\textrm{CM}(\Delta \varphi)$ oscillates with $\Delta \varphi$ \cite{Xie2012_interferometry, Xie2013, Xie2016}, see green squares in Fig.~\ref{fig:pz_asym_phase}(c). 
%
However, Fig.~\ref{fig:pz_asym_phase}(c) shows that the oscillations of $p_z^\textrm{CM}$ and $\bar{p}_z$ are almost out of phase. Thus, the $\Delta \varphi$-variation of $\bar{p}_z$ cannot be attributed to the ionization step, but may  rather be caused by the joint actions of the ZPD$_\textrm{stR}$, the 800-nm BSD and ZPD$_{\textrm{2p}\sigma_u}$ processes. 
In combination with the yield lost in the trench [see the red line Fig.~\ref{fig:pz_asym_phase}(d)], which is minimized whenever $|\bar{p}_z(\Delta \varphi)|$ becomes large, this explains the $\Delta \varphi$-dependent variation of the asymmetry, $A(\Delta \varphi, |p_z|<5)$.



\rev{But what is the reason for the observation of this trench in Fig.~\ref{fig:pz_asym_phase}(b) for $|p_z| \lesssim 1$\,a.u.? 
This trench, visible also as the dip at zero energy in Fig.~\ref{fig:pz}(c), is the signature of a suppressed dissociation probability for (near-)zero energies. 
Such suppression has been interpreted within the Floquet picture as vibrational trapping (VT) or bond-hardening on the upper LIP of the zero- or one-photon dissociation branchs, see, e.g., Refs.~\cite{Giusti-Suzor1992, Zavriyev1993, Frasinski1999, Posthumus2000, Moser2009, Badanko2016}. 
In this picture, VT can be considered the direct counter-part of the ZPD and BS processes. Other work has interpreted such suppression as the consequence of wavelength dependent weak dipole-coupling strengths of certain vibrational states \cite{McKenna2009a}. 
With the ZPD$_\textrm{stR}$ process introduced here, another possible explanation for the trapping enters the debate. 
%
%
Definite answers on the origins of trapping are beyond the scope of the current paper. However, we would like to point out that the above-discussed $\Delta \varphi$-dependent modulation of the trapped yield that shows maxima when the center of the momentum distributions overlap with the trench (e.g. at $\Delta \varphi= 0$ or $\pi$) [cf. red dots in Fig.~\ref{fig:pz_asym_phase}(d)], may be exploited by future work to obtain further insight into the dynamics leading to dissociation suppression.
}

In conclusion, we show experimentally that in the dissociation of H$_2^+$ a stimulated Raman scattering-based zero photon dissociation process, ZPD$_\textrm{stR}$, becomes active in the (near-)zero proton energy region whenever the bandwidth of the laser light is sufficient to cover the energy gap to the dissociation threshold.
This ZPD$_\textrm{stR}$ process introduced here explains the strong proton yield-enhancement in the near-zero energy region observed in many two-color experiments and also in experiments \cite{Xu2013, Kling2013a} and simulations \cite{McKenna2012a} with broadband few-cycle laser pulses centered around 730-750\,nm. 
We furthermore show that the laser field-induced asymmetry of H$_2^+$ dissociation in the (near-)zero energy region is due to the combined action of several processes rather than only pathway interferences, with bond-hardening taking a particularly important role. This finding opens up possibilities for detailed investigations of vibrational trapping and the influence of rotational states in molecular dynamics \cite{Halasz2011, Natan2016, Badanko2016}.


\acknowledgments
This work was financed by the Austrian Science Fund (FWF), Grants No. 
P21463-N22,  
P25615-N27,  
P28475-N27,  
and P30465-N27.  

\bibliographystyle{apsrev4-1}

%

\end{document}